\documentclass[11pt,twoside]{article}

\usepackage{asp2004}
\usepackage{epsf}
\usepackage{psfig}
\usepackage{lscape}

\begin{document}
\title{The {\it SPITZER} IRS view of stellar populations in Virgo early type galaxies.}
\author{A. Bressan\altaffilmark{1,2,3}, 
P. Panuzzo\altaffilmark{1},
L. Buson\altaffilmark{1}, 
M. Clemens\altaffilmark{1}, 
G.L. Granato\altaffilmark{1,2}, 
R. Rampazzo\altaffilmark{1},
L. Silva\altaffilmark{4}, 
J.R. Valdes\altaffilmark{3},
O. Vega\altaffilmark{3},
L. Danese\altaffilmark{2}} 
\affil{$^{1}$ INAF Osservatorio Astronomico di Padova, vicolo dell'Osservatorio 5, 35122 Padova, Italy}
\affil{$^{2}$ SISSA, via Beirut 4, 34014, Trieste, Italy}
\affil{$^{3}$ INAOE, Luis Enrique Erro 1, 72840, Tonantzintla, Puebla, Mexico}
\affil{$^{4}$ INAF Osservatorio Astronomico di Trieste, Via Tiepolo 11, I-34131 Trieste, Italy}

\begin{abstract} 
We have obtained high S/N {\it Spitzer} IRS spectra of 17 Virgo early-type galaxies
that lie on the colour-magnitude relation of passively evolving galaxies in the cluster.
To flux calibrate these extended sources we have devised a new procedure
that allows us to obtain the intrinsic spectral energy distribution
and to disentangle resolved and unresolved emission within the same object.
Thirteen objects of the sample (76\%) 
show a pronounced broad silicate feature (above 10$\mu$m)
which is spatially extended and likely of stellar origin,
in agreement with model predictions.
The other 4 objects (24\%) are characterized by
different levels of activity. In NGC 4486 (M87)
the line emission and the broad silicate
emission are evidently unresolved and, given also the typical 
shape of the continuum, they likely originate in the 
nuclear torus. NGC 4636 show emission lines superimposed to
extended silicate emission (i.e. likely of stellar origin, pushing the percentage of
galaxies  with silicate emission to 82\%).
Finally NGC 4550 and NGC 4435 are characterized by PAH and line emission,
arising from a central unresolved region.
\end{abstract}



\section{Introduction}
We suggested that the presence of dusty circumstellar envelopes
around asymptotic giant branch(AGB) stars should leave a signature,
a clear excess at 10 ${\mu}$m, in the mid infrared (MIR) 
spectral region of passively 
evolving stellar systems Bressan et al. (1998). 
Early detections of such an excess were suspected in M32 (Impey et al. 1986) 
from ground-based observations, and in a few ellipticals observed with ISOCAM
(Bregman et al. 1998). The first unambiguous confirmation, though barely resolved,
was found with the ISO CVF  spectrum of NGC 1399 (Bressan et al. 2001). 
Since AGB stars are luminous tracers of intermediate age and  old stellar populations,
an accurate analysis of this feature has been suggested as a complementary way to
break the degeneracy between age and metallicity that affects optical observations 
of early type galaxies (Bressan et al. 1998, 2001).
In fact the models show that a degeneracy between metallicity and age 
persists even in 
the MIR, since both age and metallicity affect mass-loss 
and evolutionary lifetimes on the AGB. But while in the optical
age and metallicity need to be anti-correlated to maintain a 
feature (colour or narrow band index) unchanged,
in the MIR it is the opposite: the larger dust-mass loss
of a higher metallicity SSP must be balanced by  
its older age. 
Thus a detailed comparison of the MIR and optical spectra of passively evolving systems
constitutes perhaps one of the cleanest ways to remove the degeneracy. Besides this
simple motivation and all other aspects connected with the detection of evolved
mass-losing
stars  in passively evolving systems (e.g. Athey et al. 2002, Temi et al. 2005), a deep look into the mid
infrared region may reveal even tiny amounts of activity (e.g. Kaneda et al. 2005).
Here we report on the first clear detection of extended silicate features in
a sample of Virgo cluster early type galaxies (Bressan et al 2006), 
observed with Spitzer IRS
instrument \footnote{The IRS was a
collaborative venture between Cornell University and Ball Aerospace
Corporation, funded by NASA through the Jet Propulsion Laboratory and
the Ames Research Center} (Houck et al. 2004) of the {\it
Spitzer Space Telescope} (Werner et al. 2004).

\begin{table}
\centering
\caption{Virgo galaxies observed with IRS}
\label{tab1}
\scriptsize
\begin{tabular}{lcccccc}
\hline
\hline
Name & V$_T$ & Date  & SL1/2 & LL2  & S/N \\
     &       &       & 60s   & 120s & 6$\mu$m. \\
\hline
NGC~4339  & 11.40  &  Jun 06 2005  & 20 &  14  &      39      \\
NGC~4365  &  9.62  &  Jan 10 2005  &  3 &  3   &      57      \\
NGC~4371  & 10.79  &  Jun 01 2005  &  9 &  10  &      40     \\
NGC~4377  & 11.88  &  Jun 01 2005  & 12 &   8  &      54      \\
NGC~4382  &  9.09  &  Jul 07 2005  &  3 &   3  &      59      \\
NGC~4435  & 10.66  &  Jun 01 2005  &  3 &   5  &      35      \\
NGC~4442  & 10.30  &  Jan 10 2005  &  3 &   3  &      46      \\
NGC~4473  & 10.06  &  Jun 01 2005  &  3 &   3  &      55      \\
NGC~4474  & 11.50  &  Jun 01 2005  & 20 &  14  &      38      \\
NGC~4486  &  8.62  &  Jun 03 2005  &  3 &   3  &      80      \\
NGC~4550  & 11.50  &  Jun 03 2005  & 20 &  14  &      42      \\
NGC~4551  & 11.86  &  Jun 03 2005  & 20 &  14  &      47      \\
NGC~4564  & 11.12  &  Jun 07 2005  &  4 &   6  &      51      \\
NGC~4570  & 10.90  &  Jun 06 2005  &  3 &   5  &      42      \\
NGC~4621  &  9.81  &  Jan 12 2005  &  3 &   3  &      63      \\
NGC~4636  &  9.49  &  Jul 08 2005  &  3 &   5  &      30      \\
NGC~4660  & 11.11  &  Jan 11 2005  &  3 &   5  &      40      \\
\hline
\end{tabular}
\end{table}

\section{Observations and data reduction}
Standard Staring mode short (SL1 and SL2) and long (LL2) low resolution IRS 
spectral
observations of 18 early type galaxies,
were obtained in the first Cycle between January and July 2005.
The galaxies were selected among those that define  
the colour magnitude relation of Virgo cluster (Bower, Lucy \& Ellis 1992),
whose common explanation
is in term of a sequence of passively evolving coeval objects
of decreasing metallicity.
The log of the observations, number of ramps (of 60s or 120s) and S/N reached at 6 $\mu$m 
are shown in Table~1.
The spectra were extracted within a fixed aperture (3.6"$\times$18" for SL and 
10.2"$\times$10.4" for LL)
and calibrated using our own software, 
tested versus the {\tt SMART} software package
(Higdon et al. 2004). 
Since the reduction procedure is described in detail in Bressan et al. 2006,
here we will only resume the main steps.
For SL observations, the sky background was removed 
by subtracting observations taken in different orders, 
but at the same node position.
LL segments were sky-subtracted by differencing the two nod
positions. 
Since the adopted IRS pipeline (version S12)
is specifically designed for
point source flux extraction,
we have devised a new procedure to flux calibrate the spectra, that exploits
the large degree of symmetry that characterizes 
the light distribution in early type galaxies.
We first obtained the real DN/Jy 
by applying 
the corrections for  aperture losses (ALCF) and slit losses (SLCF)
to the flux conversion tables provided by the Spitzer Science Center
(e.g. Kennicutt et al. 2003).
After the ALCF and SLCF corrections were applied 
we obtained the flux {\sl received} by the slit, within a given aperture
in the following way.
For each galaxy we assumed a wavelength dependent bidimensional intrinsic 
surface brightness profile and, by convolving it with the 
instrumental point spread function (PSF), we simulated the corresponding 
observed linear profile
along the slits, taking into account the relative 
position angles of the slits and the galaxy.
The adopted profile is a two dimensional
modified King law (Elson at al. 1987),
with axial ratios taken from the literature
\begin{equation}
I \equiv I_0/[1+\frac{X^2}{R_C^2}+\frac{Y^2}{(R_C \times b/a )^2}]^{\gamma/2}
\label{elson}
\end{equation}
In equation \ref{elson}, X and Y are the coordinates along the major and minor axis of the galaxies
and I$_0$, R$_C$ and $\gamma$ 
are free parameters that are 
functions of wavelength, and are obtained by fitting the 
observations with the simulated profile. 
In order to get 
an accurate determination of the parameters of the profiles, several
wavelength bins have been folded together. 
This procedure has a twofold advantage because,
it allows us both to reconstruct the intrinsic profile
and the corresponding SED, and  
to recognise whether a particular feature is resolved 
or not. Finally the spectrum, extracted in a fixed width around
the maximum intensity, is corrected by the ratio 
between the intrinsic and observed profile.
Since for the LL segment the above procedure is quite unstable,
we have preferred to fix  one of the parameters
of the profile (usually R$_c$) to the value derived
in the nearby wavelength region of the SL segment.
Finally, the LL spectrum has been  rescaled to match that of SL,
to account for the different extraction area.
\begin{figure}
\centerline{\resizebox{0.9\textwidth}{!}{
\psfig{figure=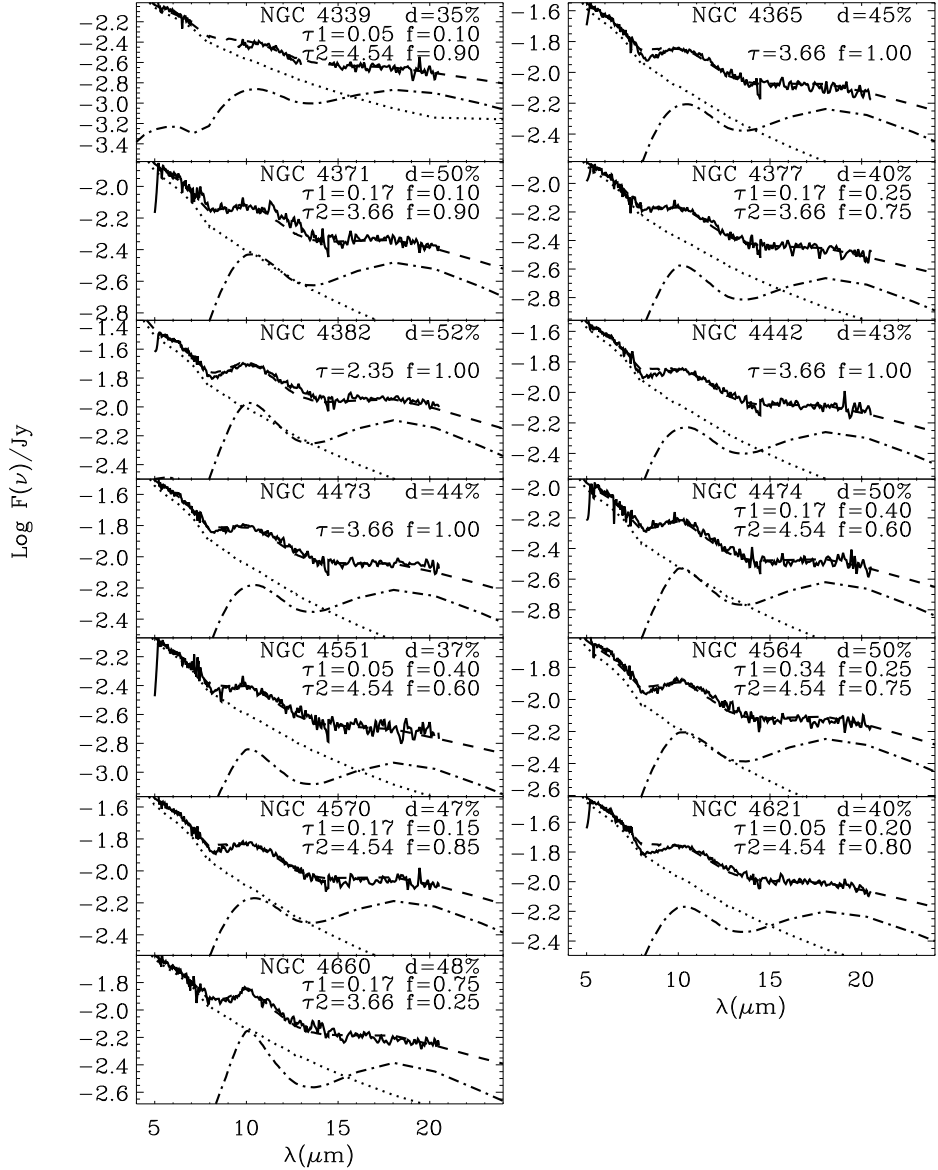} }}
\vskip 15pt
\caption{IRS spectra (solid lines) of
passively evolving early-type galaxies in the Virgo cluster.
Superimposed are best fits (dashed lines) obtained by means of simple models 
composed of an M giant star spectrum (dotted lines) from
(MARCS, Gustafsson et al. 2002) 
and a dusty silicate circumstellar envelope (dot-dashed lines) 
from Bressan et al. 1998.
Data and models are normalized at 5.5$\mu$m.
The fractional contribution of the circumstellar envelope at 10$\mu$m ("d")
and its optical depth at 1$\mu$m ("$\tau$") are indicated in the figure. 
In those cases where a combination of  two dusty envelopes is needed, 
their optical depth at 1$\mu$m and relative fractions ("f") are specified. 
}
\label{passive}
\end{figure}

\begin{figure}
\centerline{\resizebox{0.9\textwidth}{!}{
\psfig{figure=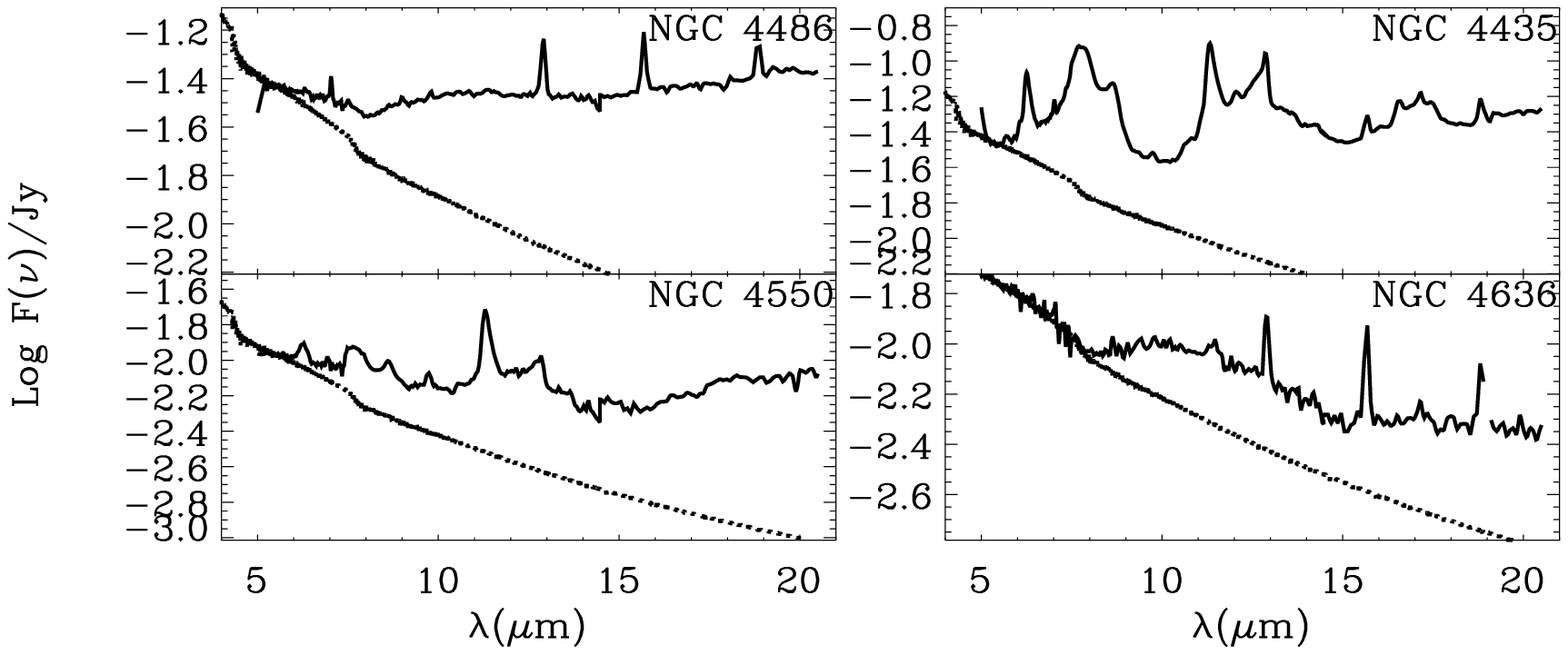,clip=} 
}}
\caption{IRS spectra (solid lines) of
active early-type galaxies in the Virgo cluster.
In this case, only the normalized M giant star spectrum 
is shown.
}
\label{active}
\end{figure}
\section{Results}

\vskip 5pt
\noindent{\sl Silicate emission from evolved stars.}

\noindent In Figure \ref{passive} we have collected  
all the galaxies whose MIR spectra are characterised 
by the presence of the broad silicate emission features 
at 10$\mu$m. 
These thirteen galaxies constitute 76\% of the sample.
The analysis of the IRS spectra indicates that the 10$\mu$m feature has 
an extended spatial distribution, consistent with that obtained in the 
spectral range dominated by stellar photospheres (below 8$\mu$m).
This is in agreement with ISOCAM observations  
that indicated spatially resolved emission at 6.7 and 15 $\mu$m
(Athey et al. 2002, Ferrari et al 2002, Xilouris et al. 2004).
The IRS spectra are very well fitted by a combination of an M giant model
spectrum (MARCS, Gustafsson et al. 2002) and dust emission from circumstellar
envelope models computed by Bressan et al. (1998) for a silicate grain
composition, as shown in Figure \ref{passive}. 
The percentage contribution of the dusty envelope to the total flux at 10$\mu$m
is roughly 50\%. This fraction is consistent with a population 
of dust enshrouded evolved stars because in spite of their paucity,
their 10$\mu$m luminosity is about 100 times the average 10$\mu$m 
luminosity of the horizontal branch  and red giant branch stars,
making their contribution to the integrated light  comparable.
{\sl If, as we believe, the 10$\mu$m feature is of stellar origin,
it constitutes perhaps the most important mid infrared
 diagnostic tool for the study of the stellar populations in 
passively evolving systems of the local universe.}
We notice however that the fits obtained here are
better than those obtained by making
use of the full isochrones (see Bressan et al. 2006). In particular 
it is surprising that the observations suggest that the major contribution comes
from envelopes in a quite narrow range of optical depth ($\tau_{1{\mu}m}$ $\simeq 3-5$),
at variance with isochrones that account for a distribution of envelopes of varying
optical depth.
This indicates that a revision  (and understanding) of the isochrones 
in the advanced phases is necessary.

\vskip 5pt
\noindent{\sl Active galaxies.}

\noindent The other four galaxies 
shown in Figure \ref{active}, display
various signatures of activity in the MIR spectra. 
They constitute
the 24\% of our sample.
It is worth noticing that
these galaxies 
had already been classified as active from optical
studies (AGN -M87-, Liner or transition Liner-HII), while no signatures of
activity had been detected in the former group.
Spectra of NGC 4636 and NGC 4486(M87) are dominated by emission lines
([ArII]7$\mu$m, [NeII]12.8$\mu$m,  [NeIII]15.5$\mu$ and  [SIII]18.7$\mu$m),
while those of  NGC 4550 and NGC 4435 by PAH emission 
(at 6.2, 7.7, 8.6, 11.3, 11.9, 12.7  and 16.4$\mu$m).
NGC 4435 shows also emission lines (and H2 S(3) 9.66$\mu$m, 
H2S(2) 12.3$\mu$m and H2S(1)17.04$\mu$m).
The broad continuum features at 10$\mu$m (and perhaps at 18$\mu$m) in M87 are
unresolved and their likely cause is silicate emission from the dusty torus 
(Siebenmorgen et al. 2005, Hao et al. 2005).
In NGC 4435 the continuum below $\sim$6$\mu$m looks extended, while above 
this (in particular PAHs) it looks unresolved. A preliminary interpretation shows
that an unresolved starburst is dominating the emission above $\sim$6$\mu$m, while 
at lower wavelengths the old extended stellar component dominates the continuum
(Panuzzo et al. in preparation).

\section{Conclusions}
We have obtained Spitzer mid infrared IRS spectra of early type 
galaxies selected along the
colour-magnitude relation of the Virgo cluster. 
We present a new method to reconstruct the SED of these galaxies,
that rests on a careful analysis of the spatial profile
sampled by the slits. 
In most of the galaxies (76\%) the emission looks spatially
extended and presents an
excess longward of 10$\mu$m, which is likely due to silicate emission from mass losing
evolved stars.
However, the peculiar distribution of the optical depth of the circumstellar 
envelopes derived from the fits requires a better understanding
of the corresponding evolutionary phases.
In the remaining smaller fraction (24\%) we detect signatures of activity
at different levels. 
If we exclude M87 which is noticeably an active galaxy, only
two out of 16 early type galaxies observed show PAHs, which corresponds to
a quite low fraction ($\sim$12\%) of the observed sample. 
It is premature to conclude that this fraction is representative of
the early type population in clusters of galaxies, in particular
if we consider that our investigation is limited to the 
brightest cluster members. A detailed comparison of our results
with those obtained for field galaxies will certainly cast light on the 
role of environment in the galaxy evolution process.

\acknowledgements 
This work is based on observations made with the Spitzer Space Telescope, which
is operated by the JPL, Caltech
under a contract with NASA. We thank J.D.T. Smith for helpful suggestions on the
IRS flux calibration procedure.
A. B., G.L. G. and L. S.  thank INAOE for warm hospitality.


\end{document}